\documentstyle[stwol,psfig]{article}

\newcommand{\bea}{\begin{eqnarray}}
\newcommand{\eea}{\end{eqnarray}}
\newcommand{\be}{\begin{equation}}
\newcommand{\ee}{\end{equation}}
\newcommand{\bd}{\begin{displaymath}}
\newcommand{\ed}{\end{displaymath}}
\def\aprle{\buildrel < \over {_{\sim}}}
\def\aprge{\buildrel > \over {_{\sim}}}
\def\ra{$\rightarrow$}
\def\dms{\hbox{$\Delta m^2$ }}
\def\dmot{\hbox{$\Delta m^2_{12}$ }}
\def\dmt3{\hbox{$\Delta m^2_{23}$ }}
\def\dmo3{\hbox{$\Delta m^2_{13}$ }}

\def\msun{\hbox{$\Delta m^2_{\odot}$ }}
\def\matm{\hbox{$\Delta m^2_{atm}$ }}
\def\mlsnd{\hbox{$\Delta m^2_{LSND}$ }}
\def\stq{\hbox{$\sin^2 2\theta$ }}

\def\bnue{\hbox{$\bar\nu_e$ }}  
\def\ne{\hbox{$\nu_e$ }}
\def\bnum{\hbox{$\bar\nu_{\mu}$ }}
\def\nm{\hbox{$\nu_{\mu}$ }}
\def\nt{\hbox{$\nu_{\tau}$ }}
\def\bnut{\hbox{$\bar\nu_{\tau}$ }}
\def\none{\hbox{$\nu_1$ }}
\def\ntwo{\hbox{$\nu_2$ }}
\def\n3{\hbox{$\nu_3$ }}
\def\ns{\hbox{$\nu_s$ }}
\def\bnus{\hbox{$\bar{\nu}_s$ }}

\def\bdbd{\hbox{$\beta\beta_{0\nu}-decay$}}
\def\ev{\hbox{eV$^2$}}
\begin{document}


\title{NEUTRINO MASSES AND OSCILLATIONS}

\author{Alexei Yu. Smirnov}

\address{ 
International Centre for Theoretical Physics\\
Strada Costiera 11, 34100 Trieste, Italy\\
\vskip 0.3cm
Institute for Nuclear Research, Russian Academy of Sciences, \\
117312 Moscow, Russia
}

\twocolumn[\maketitle
\abstracts{New effects related to refraction of 
neutrinos in different media are reviewed and 
implication of the effects to neutrino mass and mixing 
are  discussed. Patterns of neutrino masses and mixing implied by 
existing hints/bounds  are described. 
Recent results on neutrino mass generation 
are presented. They include neutrino masses in SO(10) GUT's  
and models with anomalous $U(1)$, generation of neutrino 
mass via neutrino-neutralino mixing, models of   
sterile neutrino. 
}]



\section{Introduction}

\subsection{Hints}

A number of results testifies  for  non-zero neutrino masses and mixing: 

\begin{itemize}

\item Solar neutrino spectroscopy.

\item Results on atmospheric neutrinos. 

\item Large scale structure of the Universe. 
(Its formation may imply some amount of the 
hot dark matter (HDM)).

\item LSND results. 

\item Hydrogen ionization in the Universe. 

\item Peculiar velocities of pulsars. 

\item Excess of events in tritium spectrum. 

\end{itemize}

First four  items are reviewed by  Y. Suzuki \cite{suz}, 
the fifth item  was discussed in~\cite{sciama}, 
and  the last two will be presented in sect. 
2 .

\subsection{Upper bounds}

Majority of results gives just upper bounds on 
neutrino masses and mixing. The most strong bounds relevant 
for the discussion come from:\\  
- reactor oscillation experiments BUGEY~\cite{bugey}, 
Krasnoyarsk \cite{krasn} (\bnue - $\nu_x$ oscillations );\\
- meson factory oscillation experiments (KARMEN \cite{karmen}, 
LSND \cite{lsnd}); \\
- accelerator experiments E531 (\nm \ra \nt)~\cite{E531} and 
E776 (\nm - \nt)~\cite{E776};\\  
- direct kinematic searches of neutrino mass: 
tritium experiment in Troitzk on \bnue \cite{lob}, 
(see also Maiz experiment~\cite{mainz}), 
PSI experiment on the mass of \nm  \cite{psi}, 
LEP  ALEPH result on \nt\cite{aleph}
(see also OPAL result~\cite{opal};  
new possibility to measure the mass of 
\nt has been suggested in \cite{baren}); \\
- searches for the neutrinoless double beta 
decay in  Heidelberg-Moscow experiment~\cite{heidel}, 
(see also IGEX~\cite{igex});\\
- supernova 1987A data~\cite{raf,sn87a,ssb}; dynamics of 
supernovas~\cite{raf};\\ 
- nucleosyntesis in supernovas~\cite{qian};\\
- primordial Nucleosyntesis~\cite{cris,ncris,sar};\\
- cosmology~\cite{cosm};\\
- structure formation in the Universe~\cite{struc,prim,shafi,asmi}.

These results give important restrictions on 
possible pattern of neutrino masses and mixing. 

\subsection{Lower bounds on neutrino mass?}

Neutrinos are the only fermions for which 
the Standard model  predicts masses. 
It prdicts that neutrino masses are zero. 
This follows from the content of the model, namely, 
from the fact that in the model there is  
\begin{itemize} 

\item no right handed neutrino components, 

\item no Higgs triplets which can give the Majorana mass for the 
left handed neutrinos. 

\end{itemize}

The absence of the $\nu_R$ gives an  explanation 
of strong upper bounds on the neutrino masses. 
However the  absence of $\nu_R$ looks rather anesthetic.  

The Standard Model is not the end of the story and we 
know that at least there is the gravity. The gravity can 
questioned both above items: 

1. One point is related to a consistency of the theory. 
It is argued \cite{lay}  that invariant (Pauli-Villars) 
regularization in the case of local Lorentz invariance 
requires an existence of  
16 spinors, {\it i.e.} an  additional spinor with 
properties of $\nu_R$. Once $\nu_R$ exists there is no 
reasons not 
to introduce the  Dirac mass term for  neutrinos.  

2. It is believed that  gravity breaks global 
quantum numbers. In the SM the lepton number is  global and  
therefore  one expects its violation by gravity.  
The effect of violation may  be parameterized in the 
form of the nonrenormalizable operator in the effective 
Lagrangian \cite{grav}:  
\be
\frac{1}{M_P}L L H H ~, 
\label{nonren}
\ee
where 
$M_P$ is the Planck mass, $L$ is the lepton doublet, 
$H$ is the Higgs doublet. 
The operator (\ref{nonren}) leads to the mass of neutrino
\be
m_{\nu  P} \sim 
\eta \frac{\langle H \rangle^2 }{M_P} 
\sim \frac{\eta}{G_F M_P} \sim 10^{-5} {\rm eV}~. 
\label{lmass} 
\ee 
Here $\eta$ is the renormalization group factor and   
$G_F$ is the Fermi constant.   
In fact,  the interaction (\ref{nonren}) 
allows to overcome  the problem in the second item:  
The product $H H$  plays the role of the effective Higgs triplet.

The value (\ref{lmass}) can  be considered as 
the lower bound on neutrino mass. Indeed, $M_P$ is 
the biggest 
mass scale we have in the theory. If some new interactions exist 
below this scale at  $M < M_P$,  these interactions can 
 generate 
the operator (\ref{nonren}) with $M_P$ being substituted by 
$M$. The corresponding neutrino mass,   
$\eta \langle H \rangle^2 /M$, is bigger 
than $m_{\nu P}$.  
Inverting the point, one can say that observation of 
mass $m_{\nu} > m_{\nu P}$ will testify for new physics 
below the Planck scale: 
\be
M \sim \frac{1}{G_F m_{\nu}}~. 
\label{scale}
\ee   
Note that physical scale (the scale of new particle masses, 
or condensates) can be even much smaller than the one  
estimated from (\ref{scale}). In particular,  $M$ can be a combination 
of other mass parameters $M'$, $m_{3/2}$ 
which are much smaller than 
$M$ itself: {\it e.g.} $M = (M')^2/m_{3/2}$.

A phenomenological  lower bound on 
$m_{\nu}$  has been  suggested  recently~\cite{fi}.  
The exchange of massless neutrinos leads to 
the long range neutrino forces. 
In particular, two body potential  due to the exchange of the 
$\bar{\nu} \nu$ - pair gives~\cite{fein} :
\be
V^{(2)}_0 = \frac{aG_F^2}{4\pi^3 r^5}~,
\label{zero}
\ee
where 
$a$ is known coefficient.
Many body (four,  six ... $k$ ...) potentials contain additional 
factors
$(G_F /r^2)^{2k}$ which are extremely small for $r = R_{ns}$
(radius of neutron star). 
However  
in  compact stellar objects like neutron stars and  white dwarfs, 
the contributions of 
these many body interactions to  energy of the star 
are greatly enhanced due to combinatorial factor.   

The contribution of $k$-body interactions, $W^{k}$,
to the total energy  is proportional to
number of combinations of $k$-neutrons from total number of neutrons in a 
star. The combinatorial factor leads to the series parameter
$W^{k+2}/W^{k} \sim (G_F n R_{ns})^2 \sim (10^{13})^2$, 
where $n$ is the number density of neutrons. 
So that the six body contribution to the energy dominates over
the four body contribution {\it etc}..\cite{fi}. It turns out that the energy
due to the eight body interactions
overcomes the mass of a star. According to \cite{fi} the only
way to resolve this paradox is to suggest that
all neutrinos have  nonzero masses: $m_{\nu} > 0.4 $ eV: 
Neutrino mass cuts off the forces at $r > 1/m_{\nu}$.
In section 2.3 we will argue that 
there is another resolution of the paradox and 
the mass of neutrino can be zero.

\section{Refraction and neutrino masses}

There are several  new  results on 
neutrino refraction and propagation in media  
which have important implications to the neutrino mass problem.

\subsection{Effective potentials}

In transparent medium neutrinos undergo essentially 
elastic forward scattering. The effect of the scattering 
is described by 
\be
H = 
\frac{G_F}{\sqrt{2}} \bar{\nu}\gamma^{\mu} (1 - \gamma_5) \nu 
\langle \psi_e | \bar{e}\gamma_{\mu} (g_V + g_A \gamma_5) e|\psi_e\rangle 
~, 
\label{ham}
\ee
where $\psi_e$ is the wave function of medium. 
(We took into account the interactions with electrons only).  
For ultrarelativistic neutrinos the expression 
(\ref{ham}) can be reduced to 
\be 
H = 
\sqrt{2}~ G_F~V ~{\nu}^{\dagger} \nu ~, 
\ee
where $V$ is the effective potential. 
Let us summarize the results on  the potentials 
for some  cases: 

1. Unpolarized medium at the rest: Only 
$\gamma^0$ component of the vector current contribute to $V$ 
and its matrix element gives the density of electrons, $n_e$. As the 
result we get~\cite{wolf}
: 
\be 
V  = \sqrt{2} G_F n_e g_V~. 
\ee

2. Polarized medium at the  rest. 
The axial vector current, $\vec{\gamma} \gamma_5$, 
also gives the contribution which is proportional 
to the vector of spin~\cite{lang}
: 
\be 
V  = \sqrt{2} G_F n_e \left[ g_V + g_A 2 (\vec{k} \cdot 
\vec{\langle s \rangle}) \right]~, 
\label{polarm}
\ee
where  $\vec{k} \equiv \vec{p}/p$, and   
$\vec{p}$ is the momentum of neutrino, 
$\vec{\langle s \rangle}$ 
is the averaged spin of electrons in medium. 
The second term can be rewritten as 
$\sqrt{2} G_F g_A (n_+ - n_-)$.   
Here  
$n_+$,  $n_- $ are the concentrations of the electrons with 
polarization along and against the neutrino momentum.    

3. In the case of moving medium also spatial components 
of the vector current give non-zero contribution: 
$
\langle \psi_e | \vec{\gamma}|\psi_e\rangle \propto \vec{v}
$ 
and \cite{prog}
\be 
V  = \sqrt{2} G_F n_e g_V ( 1 - v\cdot \cos \theta)~, 
\ee
where $\theta$ is the angle between the momentum of the electrons
and neutrino. 
In the case of isotropic distribution the correction disappears. 
In this case non zero effect of the motion appears  via   the 
correction to the propagator of the vector boson: 
$G_F \rightarrow G_F (1 + q^2 /m_W^2)$, 
where $q^2$ is the four momentum of the intermediate  boson squared 
\cite{prog}. 
In thermal bath $q^2 \sim T^2$ and 
one gets \cite{notz}
\be 
V  \sim  \sqrt{2} G_F n_e  A \frac{T^2}{m_W^2} ~, 
\label{term}
\ee
where $A$ is the constant which depends on the composition of plasma. 
In all the cases, apart from 
the thermal correction (\ref{term}),   
$V$ has  opposite signs for neutrinos and antineutrinos.

\subsection{Neutrino sea and the long range neutrino forces}

At low energies a medium is transparent for neutrinos
and main effect is the refraction. 
Refraction index equals:
\be
(n_r - 1) = 
\frac{V}{p} \propto \frac{G_F n}{p}~.
\ee
At usual conditions: $E \sim 1$ MeV, $\rho = 1 $ g/cm$^3$,  the
deviation of the refraction index from 1 is extremely
small: $(n_r - 1) \sim 10^{-20}  - 10^{-19}$. However at very low
energies this deviation can be of the order one,
leading to complete inner reflection of neutrinos in  stars \cite{loeb}.
For neutron star with $\rho \sim 10^{14}$ g/cm$^3$ the complete
reflection takes place for neutrinos with  energies $E < 50$ eV.
In other terms,
a star can be considered as a potential well with the depth
$V$. The potential has different signs for neutrinos and antineutrinos.
Therefore,  neutrinos are trapped, whereas antineutrinos are
expelled from the star. In such a way strongly degenerate sea is formed with
chemical potential \cite{loeb}
\be
\mu \sim V \sim G_F n~.
\ee
In neutron stars the density of  neutrinos from the sea is
$n_{\nu} \sim 10^{17}$ cm$^{-3}$ and the total energy in the sea is very
small in comparison
with mass of a star. In spite of this, an existence of the sea can play an
important role.  
The degenerate sea in stars leads to  Pauli blocking
of the long range forces. Instead of (\ref{zero}) we get for two body
potential \cite{fr}:
\be
V^{(2)}_{\mu} = \frac{aG_F^2}{4\pi^3 r^5} (\cos 2\mu r + \mu r \sin 2\mu
r).  \label{pauli}
\ee
Note that $1/\mu \sim 10^{-5}$ cm $ \ll R_{ns}$. Rapidly oscillating
factors in (\ref{pauli}) lead to effective cut off of the forces at
$r > 1/\mu$. Similar oscillating factors  appear for many body
interactions. As the result the many body forces do not dominate in self
energy
of star. This can resolve the energy paradox suggested in~\cite{fi} 
even for massless neutrinos \cite{fr}.

Another objection to Fischbach
result is related to
resummation of series over the $k$-body interactions~\cite{gav}.
The interaction with medium modifies the dispersion relation
for neutrinos:
\be   
q_0 = |\vec {q}| \pm V~, 
\ee
and correspondingly, the propagator of neutrino:
\be
S(q) = \frac{i}{\hat{q}_v} =  
\frac{i}{(q_0 - V) \gamma^0 - \vec{q} \vec{\gamma}}~.
\ee
This dressed propagator is the  sum
of free propagator and  the results of elastic forward scattering on one
neutron, two neutrons .... $k \rightarrow \infty$ neutrons in medium.
If the neutrino forms closed loop, then this process is equivalent to
summation of 0, 2,  4, .... k body interactions due to
neutrino exchange. Therefore the energy density due to
the neutrino exchange
can be written as
\be
w = \int \frac{d^4 q}{(2\pi)^4} (-i)
Tr\left[ q^0 \gamma^0 \frac{1}{\hat{q}_v} \frac{1 - \gamma_5}{2}
\right] ~.
\label{energy}
\ee 
The energy density due to 
the interactions,  $\Delta w$,  
is the difference of $w$ given in (\ref{energy}) and  $w_0$ - the energy 
density for vacuum propagator: 
$\Delta w = w - w_0$.
Total energy of star is the integral
of $\Delta w$ over the volume of  star.
In approximation of  uniform medium, $V = const$,
one can redefine the integration variable in  (\ref{energy}):
\be
q'_0 = q_0 - V ~.
\label{redef}
\ee
After redefinition $w$ is reduced to $w_0$, so that
$\Delta w = 0$.
Thus the energy of a  star in this approximation is zero.
However, this proof corresponds to infinite  and uniform
medium. Real star has finite size and the distribution of
neutrons is non-uniform. In this case the redefinition of variables
(\ref{redef}) is impossible and non-zero self energy of the star appears. 


\subsection{Oscillations in Magnetized Medium}

Let us consider neutrino propagation in 
the thermal bath with magnetic field.  
Effect of the medium can be calculated as the correction to self-energy. 
Two diagrams appear: The loop diagram with $W$-boson: 
$\nu \rightarrow W e \rightarrow \nu$,  where for 
the electron  we should use the effective propagator in thermal bath. 
(ii) the tadpole diagram with $Z$ and electron in the loop. 
The electrons couple to the electromagnetic field~\cite{olivo}. 

In strongly degenerate gas, $E_F \gg T$, where 
$E_F$ is the Fermi energy one gets the following expression 
for the effective potential in the magnetic field $B$~\cite{olivo,magn}: 
\be 
V  = 
\sqrt{2} 
G_F n_e  g_V + 
\frac{e g_A G_F}{\sqrt{2}}
\left(\frac{3 n_e}{\pi^4} \right)^{\frac{1}{2}}    
( \vec{k} \cdot \vec{B})~. 
\label{vmag}
\ee
The correction originates from the axial vector current. 
It  influences   dynamics of the neutrino conversion.  
In particular, the correction  modifies the resonance condition: 
\be
V  + \frac{\Delta m^2}{2E} \cos2\theta = 0
\ee 
shifting position of the resonance in comparison with 
the case of zero magnetic field. It also influences 
the adiabaticity condition. 

There are however wrong statements that the magnetic term can compensate 
or  even be bigger that 
the first (vector current) term. It would induce new resonances 
and open the possibility to have the flavor resonances both for 
neutrinos and antineutrinos 
in the same medium. 
Actually magnetic (axial) term can not  be bigger  than 
the vector one. 
This can be seen immediately from another approach 
to the problem~\cite{nuno}. 

Indeed,  the effect of the magnetic field is reduced to 
polarization of electrons,  so that one can use the result 
(\ref{polarm}) 
for the effective potential and calculate 
the average polarization of the electrons. 
For flavor oscillations the matter effects is determined by 
charge current scattering on electrons 
for which $g_A = g_V = 1$ and therefore 
\be 
V  = \sqrt{2} G_F n_e \left(1 +  
2 \langle s \rangle \cos\alpha \right)~. 
\ee
Here $\alpha$ is the angle between the 
neutrino momentum and the polarization of  electrons  
and ${\langle \vec{s} \rangle} = {\langle \vec{s} ( B) \rangle}$. 
Obviously, second term can not be bigger than 1,  
so that one can get at most the compensation of the 
effective potential: $V = 0$ in the case of the complete 
polarization of electrons in the direction against the neutrino 
momentum. Complete polarization can be achieved in the case of 
very big magnetic field  and zero temperature. 

The polarization equals $(n_+ - n_-)/n_e$,   
where $n_+$, $n_-$:
are the concentrations  of the electrons with 
polarization + 1  and  - 1. 
The energy spectrum of electrons in the magnetic field 
is quantized: 
\begin{equation}
\varepsilon (p_z,n,\lambda)=\sqrt{p_z^2 + m_e^2 +
\mid e\mid B(2n + 1 - \lambda)} ~, 
\label{spectrum}
\end{equation} 
where $\lambda = - 2s_z$. 
It consists of 
main Landau level, $n = 0$, $\lambda = 1$, and pairs of the 
degenerate levels with opposite polarizations. 
Therefore the polarization effect is determined by 
concentration of electrons  in 
Landau level, 
\be
2 \langle s \rangle = \frac{n_+ - n_-}{n_e} = \frac{n_0}{n_e}~.  
\ee
For strongly degenerate gas:  
\be
n_0 =  \frac{eBp_{F}}{2\pi^2} ~, 
\ee 
where the Fermi momentum, $p_F$,  is determined by 
the normalization \cite{sem} 
\begin{equation}
n_e = 
\frac{eBp_{F}}{2\pi^2} 
+ \sum_{n =
1}^{n_{max}}\frac{2eB\sqrt{p_{F}^2 - 2eBn}}{2\pi^2}~. 
\label{dens1}
\end{equation} 
The  first term corresponds to  the main Landau level
$n = 0,~ \lambda = 1$: 
and the second one  
is the result of summation over all other levels. 
The complete polarization corresponds to 
$2eB\geq p_{F_e}^2$, when  
$ n_{max}\leq 1$, and the 
sum  vanishes.  
In this case  all electrons are in  the main Landau
level:   
$n_e = eBp_{F}/ 2\pi^2 $, from this one gets   
$p_F  = 2\pi^2 n_e/eB$,  
and consequently,  $n_0 = n_e$. 
In the limit of small field:   
$p_F \approx (3\pi^2 n_e)^{1/3}$  and 
\be
n_0 =  \frac{eB }{2}\frac{3 n_e}{\pi^4}~.  
\ee 
This leads to the result (\ref{vmag}). 

For oscillation to sterile neutrinos, however,  the effective 
$g_A$ can be bigger than $g_V$ and the level crossing 
phenomena induced by magnetization are possible \cite{nuno}.  

\subsection{Neutrino mass and the peculiar velocities of pulsars}

Important application of results described in sect. 2.3. 
has been found by Kusenko and Segre~\cite{kus}. 
There is the long standing problem of explanation of 
the high  peculiar velocities of pulsars 
($v \sim 500$ km/s). Non-symmetric collapse, effects 
in binary systems {\it etc.} , give typically smaller 
velocities.   
 
It looks quite reasonable  to relate 
these velocities with neutrino burst~\cite{puls}. 
The momenta   of pulsars are  
$10^{-3} - 10^{-2}$ 
of the integral momentum  
carried  by neutrinos. Therefore, 
$10^{-3} - 10^{-2}$ 
asymmetry (anisotropy) in neutrino emission is enough 
for explanation of the peculiar velocities~\cite{puls}. 

The anisotropy of neutrino properties  can 
be related to the magnetic field. It was suggested that 
very strong magnetic field ($10^{15} - 10^{16}$ Gauss) 
can influence the weak processes immediately: the probability 
of emission of neutrino along the field and against the field 
are different.  

According to  mechanism suggested  in \cite{kus}  
magnetic field influences the resonance flavor 
conversion leading to angular asymmetry of the conversion 
with respect to the magnetic field. The latter 
results in  asymmetry of the neutrino properties.

It is assumed that the resonance layer for the 
conversion 
$\ne - \nt$ lies between  the \ne-neutrinoshpere and 
\nt-neutrinospere (the latter is deeper than the former 
due to weaker interactions of \nt). Thus the \nt which 
appear in the 
resonance layer will propagate freely and \ne are immediately 
absorbed. The resonance layer becomes the ``neutrinosphere" for \nt.  
(In fact, in presence of the magnetic field the neutrinosphere 
becomes ``neutrinoellipsoid" and this is crucial for the mechanism).  

It is assumed that inside the  protoneutron star there is 
a strong magnetic field of the dipole type. 
Then in one semishpere the field is directed outside the star,   
so that for neutrinos leaving the star 
$(\vec{k} \cdot \vec{B}) > 0 $, 
whereas in another semisphere the field  points towards  the center of 
star    and 
$(\vec{k} \cdot \vec{B}) < 0$ . 
Since the electronic gas in the star is strongly degenerate 
we can use the expression (\ref{vmag}) for the effective potential.  
According to (\ref{vmag}) the magnetic 
field modifies the resonance condition differently in 
these two semispheres. 
In semishpere with $(\vec{k} \cdot \vec{B}) < 0 $, 
the resonance condition is satisfied at larger densities 
and  larger temperatures;  \nt emitted 
from this semisphere will have bigger energies. 
On the contrary, in the  neutrinosphere
with $(\vec{k} \cdot \vec{B}) > 0$ the resonance is 
at lower densities and lower  temperatures  and  
neutrinos have smaller energies.  Thus presence of the magnetic field 
leads to difference in energies of \nt  emitted in different 
directions and therefore neutrino burst knocks  the star. 
The observed velocities imply the polarization effect 
 $10^{-3} - 10^{-2}$, or according to 
(\ref{vmag}) 
$$
eB \left(\frac{3 n_e}{\pi^4} \right)^{\frac{1}{3}} \sim 
10^{-3} - 10^{-2}~.     
$$
Below the \ne - neutrinosphere: 
$n_e > 10^{11}$ cm$^{-3}$ which gives 
$B \sim 10^{13}$  Gauss.   
From the condition that the resonance should be 
below the \ne -  neutrinosphere one gets 
\be 
\Delta m^2 > 10^4~~ {\rm eV}^2~,~ 
{\rm or} ~~ m_3 > 100 ~~ {\rm eV}~.
\label{puls}
\ee  
The mixing angle can be  rather small: 
from the adiabatic condition it follows  
$\stq > 10^{-8}$.  

Thus explanation of the peculiar velocities of  
pulsars based on the resonance flavor conversion 
implies the mass of the heaviest ($\sim$ \nt) neutrino 
bigger than $100$ eV. 
To avoid the cosmological bound on mass,  the neutrino  
must decay (e.g. with Majoron emission).  
The attempts to diminish $m_3$ by means  of very large magnetic 
field (so that the polarization effect compensates the 
density) lead to very strong asymmetry $\sim 1$. Another problem is that  
due to relatively high temperatures 
very strong polarization and consequently, the 
compensation are impossible.  

In  connection with Kusenko-Segre proposal 
it is interesting to mark 
recent results on measurements of the 
beta spectrum in tritium decay~\cite{lob}. 
There are two features in the spectrum: 
(i) Excess of events near the end point, $Q$, of the  spectrum 
$Q - E_e \aprle 10$ eV,  
(peak in the differential spectrum) which leads 
to the negative value of the $m^2$ in 
usual fit. 
(ii). Excess of events at lower energies of the electrons: 
$Q - E_e \aprge 200$ eV. The excess in this region was also 
observed by  Mainz group.  
One possible explanation of this anomaly is an existence 
of  neutrino  with mass $m \sim 200$ eV whose 
admixture in the 
electron neutrino state 
is characterized by probability 
$P \sim 1 - 2$ \% . This is precisely in the range implied by 
pulsar velocities. 

As far as  the first anomaly is 
concerned (the negative 
$m^2$) one possible  explanation is the tachionic nature of 
neutrinos \cite{cib}. It should be stressed, however, that position of the 
peak depends on condition of the experiment: 
In the run of  experiment in  1994  the peak was at 
$Q - E_e \approx 7$ eV 
whereas in the run 1996 the peak is at 
$Q - E_e \sim 11$ eV. 
There were some changes of the experiment in run 1996, in particular, 
the strength of the magnetic field was higher. 
The shift of the peak indicates that 
it may have the instrumental origin, 
rather then the origin in  neutrino properties.   

\subsection{Lepton asymmetry in the Early Universe}

According to (\ref{term}) in the Early Universe the difference 
of the potentials for different neutrino species  
can be written as 
\be 
\Delta V  = \sqrt{2} G_F n_{\gamma} 
\left( \Delta L + A \frac{T^2}{m_W^2}\right)~, 
\label{euniv}
\ee
where $n_{\gamma}$ is the photon density, 
$\Delta L = (n_L - n_{\bar{L}})/n_{\gamma}$ 
is the leptonic asymmetry and 
$n_L$,  $n_{\bar{L}}$ are the concentrations of the 
active neutrinos and antineutrinos.

Matter effects can be important for  oscillations into 
sterile neutrinos. Matter influences differently the neutrino and 
antineutrino 
channels, so that transitions 
\nt $\rightarrow$ \ns , and \bnut $\rightarrow$ \bnus 
can create the \nt - \bnut asymmetry in the Universe.

Since  $V$ depends on concentration of neutrinos 
themselves,  
and consequently,   on  conversion probability, 
the task becomes non-linear. Due to this,  
depending on values of parameters, a  
small original asymmetry (one can expect 
$\Delta L_0 \sim \Delta B \sim 10^{-9}$) can be  
further suppressed \cite{tign} or blow up \cite{foot,thom}. 
The leptonic asymmetry  influences 
the primordial nucleosynthesis. 
It was realized recently, 
that  it can suppress production of  sterile 
neutrinos, so that the concentration of these neutrinos 
is much smaller than the equilibrium concentration 
even in the case  of large mixing angle 
and large mass squared difference. 

Scenario suggested in \cite{foot,thom} is the following. 
Suppose \nt mixes with  \ns and  parameters of the system are: 
\dms $\sim 5$ \ev ~ and  
$\theta_{\tau s} \sim 10^{-4}$. 
On the contrary,  \nm - \ns   
has large mixing
$\theta_{\mu s} \sim 1$ and \dms $\sim 10^{-2}$ \ev , 
so that \nm - \ns oscillations can solve the atmospheric 
neutrino problem.   It turns out that in spite of 
this large mixing the concentration of 
sterile neutrinos is small. 

Let us consider the evolution of system with 
decrease of temperature. 
There are two important scales  determined by the equality 
of the $T$-term in  $\Delta V$   and  level splitting due to mass 
difference: 
\be 
\sqrt{2} G_F n_{\gamma} A \frac{T^2}{m_W^2} = 
\frac{\Delta m^2}{2T}~.  
\label{temp}
\ee
For 
\dms corresponding to \nt - \ns and \nm - \ns channels 
we get from (\ref{temp}) 
$T_{\tau} 
\approx 14$ MeV and 
$T_{\mu} 
\approx 2$ MeV. 
(i) 
For $T > T_{\tau}$ the oscillation transitions are suppressed by 
$T$ term in the potential.  
(ii)  For $T \sim T_{\tau}$ the $T$-term drops enough and oscillations 
become possible. Due to non linearity of the equations  
the amplitude of oscillations blows up and the 
asymmetry  reaches (practically during the same epoch $t \sim 10^{-2}$ 
sec)    $\Delta L \sim 10^{-5}$. With further diminishing of temperature 
the asymmetry may  slowly increase up to $10^{-2}$ 
or even higher.   (The 
mixing is chosen to be small enough,  so that 
the concentration of sterile neutrinos,  
$n_s \aprge n_{\gamma} \Delta L$,  is still smaller than the 
equilibrium one). 
(iii) In the epoch 
$T \aprle T_{\mu}$, when transition \nm - \ns 
could be important, the effective 
(matter) mixing \nm - \nt is 
suppressed by leptonic asymmetry ($\Delta L$ -term of the potential) 
produced previously in \nt oscillations.  We discuss the 
application of this result in sect. 4.7. 


\section{Pattern of neutrino masses and mixing}

\subsection{Neutrino anomalies}

Existing neutrino anomalies imply strongly different 
scales of \dms . 
For the solar neutrinos,  the atmospheric neutrinos and  
LSND we have correspondingly:  
\be
\Delta m_{\odot}^2 \sim (0.3 - 1.2)\cdot 10^{-5}~ {\rm eV}^2 ~, 
\ee
\be
\Delta m_{atm}^2 \sim (0.3 - 3)\cdot 10^{-2}~ {\rm eV}^2~,  
\ee
\be
\Delta m_{LSND}^2 \sim (0.2 -  2)~ {\rm eV}^2 ~.  
\ee
That is 
\be
\Delta m_{LSND}^2 \gg
\Delta m_{atm}^2 \gg
\Delta m_{\odot}^2. 
\label{hierarchy}
\ee
The  mass scale which gives desired HDM component 
of the Universe, $m_{HDM}$:  
\be
m_{HDM}^2 \sim  (1 - 50)~ {\rm eV}^2 
\ee
can cover the LSND range. 

In the case of three neutrinos 
there is an obvious relation: 
\be
\Delta m_{21}^2 + 
\Delta m_{32}^2 =
\Delta m_{31}^2. 
\label{nasseq}
\ee
and inequality (\ref{hierarchy}) can not be satisfied. 
That is with three neutrinos it is impossible to reconcile 
all the anomalies.  Furthermore, 
additional  bigger scale is needed for explanation of the 
pulsar velocities (\ref{puls}). 

Three different possibilities are discussed  in this connection. 
One can 
\begin{itemize}
\item suggest (stretching the data) that 
\be
\Delta m_{LSND}^2 = 
\Delta m_{atm}^2 ~;
\label{full}
\ee 
Also the possibility $\Delta m_{\odot}^2 = 
\Delta m_{atm}^2$ was discussed~\cite{anjkr}

\item ``sacrifice" at least one anomaly, {\it e.g.} the LSND result, 
or atmospheric neutrinos;

\item
introduce additional neutrino states. 
\end{itemize}

In what follows we will consider examples which realize these 
three  possibilities. 

There is another important mass scale: 
the upper bound on  
the effective Majorana mass of the  electron neutrino 
which determines  the rate of the 
neutrinoless double beta decay: 
\be
m_{ee} = \sum_{i = 1,2,3} U_{ei}^2 m_i . 
\ee 
Here $U_{ei}$ are the elements of the lepton mixing matrix. 
Taking into account uncertainties in the nuclear matrix element
one gets from the data 
$m_{ee} \aprle 0.5 - 1.5 ~{\rm eV}$ \cite{}. 
Forthcoming experiments (NEMO-III \cite{nemo}) will be able to strengthen 
the bound by factor 3.  
Note that typically $m_{HDM} > m_{ee}$.

\subsection{Everything with three neutrinos?}

It is assumed that LSND and atmospheric neutrino scales 
coincide \cite{full}:
\be
\Delta m_{23}^2  = \Delta m_{LSND}^2 = 
\Delta m_{atm}^2 \sim 0.2 - 0.3~ {\rm eV}^2 ~, 
\label{stret}
\ee 
\none and \ntwo are strongly mixed in \nm and \nt.  
The neutrino \none has dominant \ne - flavor and 
weakly mixes with \ntwo . The mass splitting between these two states 
\dmot $\approx $ \msun   (see fig.1).  
Basic features of this scenario are the following:\\ 
\begin{figure}
\centerline{
\psfig{figure=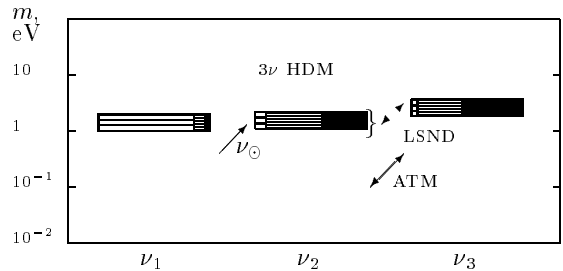,height=1.4in,angle=0}}
\caption[]
{
Qualitative pattern of the neutrino masses and mixing. 
Boxes correspond to different mass eigenstates. The sizes of different 
regions in  the boxes determine flavors (  
$|U_{if}|^2$) of given eigenstates. Weakly hatched regions 
correspond to the electron flavor (admixture of \ne), 
strongly hatched regions depict the muon flavor and black 
regions present the tau flavor.  Arrows connect the eigenstates 
involved in oscillation/conversion which solve 
$\nu_{\odot}$ - solar, ATM - atmospheric, LSND - problems.    
Scenario shown here reproduces simultaneously 
$\nu_{\odot}$, ATM, LSND and HDM.    
\label{fig:1}
}
\end{figure}
\noindent
(i) \nm - \nt oscillations explain the 
atmospheric neutrino deficit. However,  since \dmt3 is rather  big,   
no appreciable angular 
dependence is expected for  multi GeV events in Kamiokande 
and SuperKamiokande.\\  
(ii)  The solar neutrino problem is solved by 
\ne \ra  \nm resonance conversion. \\
(iii) The probability of the LSND/KARMEN oscillations 
is determined by 
\be
P \propto  4 |U_{3e}|^2 |U_{3 \mu}|^2 ~.
\ee  
(iv) The scenario can supply three component HDM,  
if the absolute values of the masses  are  
in eV range.  In this case the spectrum is    degenerate:  
$m_1 \approx m_2 \approx m_3 \approx$ 1 eV . \\
(v) If neutrinos $\nu_i$ are the Majorana particles,  
then  $m_{ee} \approx m_1 \sim 1$ eV is at the level of present 
upper experimental bound \cite{}.  

This scheme is a variant of the previously 
considered schemes with 
three degenerate neutrinos  and 
an order of magnitude smaller mass splitting: \dmt3 $\sim 10^{-2}$ 
\ev (see sect. 4.4).

 One can modify the scenario assuming mass hierarchy, so that 
$\Delta m_{23}^2 \approx m_3^2$. 
In this case $m_3 \sim 0.5$ eV,  $m_2 \approx 3 \cdot 10^{-3}$ eV 
and  $m_1 \ll m_2$. 
The contribution to HDM is small and  signal in  
\bdbd ~ searches is negligible. 


\subsection{Sacrifice solar neutrinos}

The scheme \cite{rayc} consists of   two heavy  degenerate 
neutrinos \ntwo ,  \n3  strongly mixed in 
\nm ,  \nt and one light weakly mixed state \none (fig.2):  
\begin{eqnarray}
m_1 \ll m_2 \approx m_3 \approx 1 {\rm  eV} , 
\nonumber \\
\dmt3 \approx 10^{-2} {\rm eV}^2 
\label{zees} 
\end{eqnarray}
\none $\approx$ \ne. This scenario 
is realized {\it e.g.}  in the 
Zee model. Basic features of the scenario are the following.\\  
\begin{figure}
\centerline{
\psfig{figure=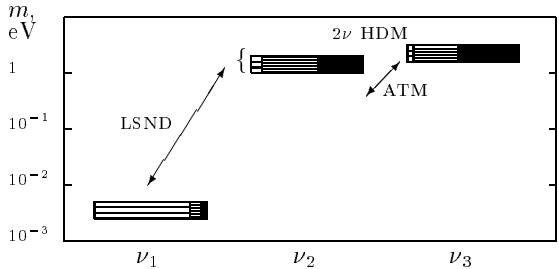,height=1.4in,angle=0}}
\caption[]{
The same as in fig.1. Scenario without explanation of 
the solar neutrino deficit.  
\label{fig:2}}
\end{figure}
\noindent
(i) Atmospheric neutrino problem is solved by 
\nm - \nt oscillations.\\  
(ii) \none and \ntwo form two components HDM.\\ 
(iii) The probability of oscillations 
in LSND/KARMEN experiments is determined by 
$e$  and $\mu$ flavors of the lightest state: 
\be
P \propto  4 |U_{e1}|^2 |U_{\mu1}|^2 ~.
\ee
Mixing elements  $U_{e1}$ and $U_{\mu1}$ 
are restricted 
 by BUGEY and BNL E776  experiments.\\ 
(iv) No observable signal of \nm - \nt oscillations 
is expected in CHORUS/NOMAD experiments \cite{chor,nom}, 
however these experiments may discover \ne - \nt oscillations.\\ 
(v) \bdbd ~~is strongly suppressed. \\ 

Modification of the scenario 
is suggested with the same mass spectrum 
but inverse flavor hierarchy 
\cite{rayc}  (fig. 3). Electron flavor is essentially in 
heavy states and its admixture in \none is small. 
Tau and $\mu$ flavors have comparable admixtures in all three states. 
Some features of the scenario are:\\ 
\begin{figure}
\centerline{
\psfig{figure=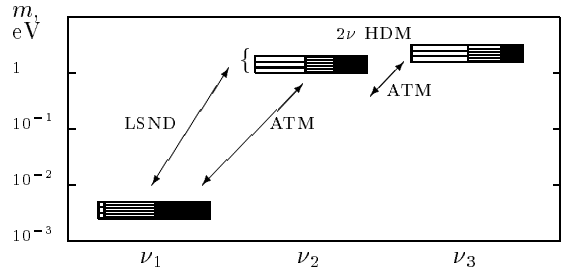,height=1.4in,angle=0}}
\caption[]{
The same as in fig.1. Scenario 
with inverse flavor/mass hierarchy and without 
explanation of 
the solar neutrino deficit.  
\label{fig:3}}
\end{figure}
\noindent
(i) All three flavors and both mass squared differences 
contribute to 
 the oscillations of  atmospheric neutrinos. 
The generic 3$\nu$-case is realized. \nm - \ne   
has unsuppressed mode of oscillations with \dmt3~, and \nm - \nt has both 
\dmot and \dmt3 modes. CHOOZ experiment \cite{chooz} will put the bound on 
this possibility.\\
(ii) Effective Majorana mass of the electron neutrino  
is $m_{ee} \approx m_0 (U_{e2}^2 + \eta_{CP} U_{e3}^2) $,  
where $\eta_{CP}$ is the relative CP-parity of two massive neutrinos.  
The bound from \bdbd  \ can be satisfied 
by some amount of cancelation.\\ 
(iii) If $m_0^2 > 4$ \ev , ~~ CHORUS/NOMAD 
may observe  signal of \nm - \nt oscillations.\\  
(iv) Due to  inverse flavor/mass 
hierarchy the scenario predicts strong 
resonance conversion of antineutrinos in supernova: 
\bnum, \bnut \ra \bnue. The conversion results in 
permutation of \bnut,  \bnue energy spectra which  
is disfavored by SN87A data \cite{ssb}. 

In these schemes the solar neutrino data  can be explained 
by virtue of 
introduction of the additional (sterile) neutrino states.

\subsection{Sacrifice LSND. Degenerate neutrinos}

Solar, atmospheric and HDM problem can be solved
simultaneously, if neutrinos have strongly
degenerate  mass spectrum 
$m_1 \approx m_2 \approx m_3 \sim 1 -2 $ eV  \cite{cald,pets,josh}, 
with $\dmot = \msun = 6 \cdot 10^{-6}$ \ev 
and $\dmt3 = \matm = 10^{-2}$ \ev (fig.4). 
\begin{figure}
\centerline{
\psfig{figure=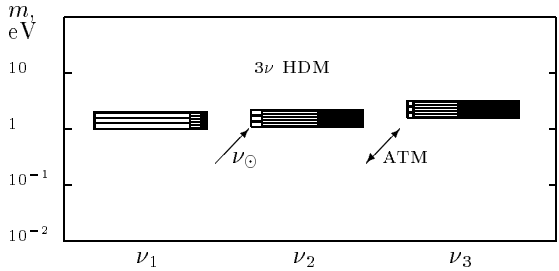,height=1.4in,angle=0}}
\caption[]{The same as in fig.1. Scenario with
strongly degenerate neutrino spectrum. 
\label{fig:4}}
\end{figure}
The corresponding mass matrix may have the form 
\cite{pets}
\begin{equation}
m = m_0 I + \delta m ,
\label{picture}
\end{equation}
where $I$ is the unit matrix,  
$\delta m \ll m_0 \approx 1 - 2$ eV.
Moreover (\ref{picture}) can be realized in  unique see-saw mechanism 
with non zero direct Majorana masses of the left components.  
Main contribution, $m_0$,  originates from  interaction
with Higgs triplets which respects some horizontal symmetry like
$SU(2)$, $S_4$ or permutation symmetry. 
It looks quite interesting  that
the desired mass splitting $\delta m$ can be generated by the
standard see-saw contribution with
$M_R \sim 10^{13}$ GeV \cite{pets}.

The effective Majorana mass  
$m_{ee} \approx m_0$ 
is at the level of upper bound 
from the \bdbd. 

The mass  
$m_{ee}$ can be suppressed~\cite{nuss}
if the electron  
flavor has large admixture in \none and \ntwo , 
so that the solar neutrino problem is solved by 
the large mixing MSW solution. 
Now the effective Majorana mass 
equals 
$m_{ee} \approx m_0 (1 - \sin^2 2\theta)$,  and for 
$\stq = 0.7$ one gets  suppression  factor 0.3. 
However simple formula (\ref{picture}) does not work   
\cite{nuss}. 

No observable signals are expected in CHORUS/NOMAD and 
LSND/KARMEN. \\

\begin{figure}
\centerline{
\psfig{figure=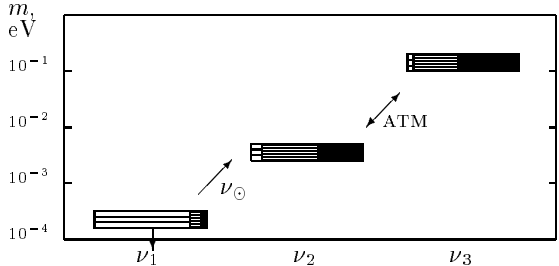,height=1.4in,angle=0}}
\caption[]{The same as in fig.1. Scenario 
for solar and atmospheric neutrinos. 
\label{fig:5}}
\end{figure}
Another possibility (fig.5)
is to sacrifice the HDM assuming (if needed) that some
other particles ({\it e.g.} sterile neutrinos, axino etc..)
are responsible for  structure formation in the Universe.
In this  case $m_3 \sim 0.1$ eV and 
$\nu_{\mu} - \nu_{\tau}$ oscillations explain  the
atmospheric neutrino deficit. Strong
\nm - \nt mixing, could be related to  relatively small mass splitting
between $m_2$ and $m_3$ which implies the enhancement of the
mixing  in the neutrino Dirac mass matrix~\cite{fty}.
It could be related to  the see-saw enhancement mechanism~\cite{sss,tan}
endowed by renormalization group
enhancement~\cite{tan} or with strong mixing in  charge lepton sector
~\cite{leon}.  

\subsection{Without the atmospheric neutrino problem}

The schemes are suggested  which can accommodate  solar 
neutrinos, HDM, and the  LSND result.  
\begin{figure}
\centerline{
\psfig{figure=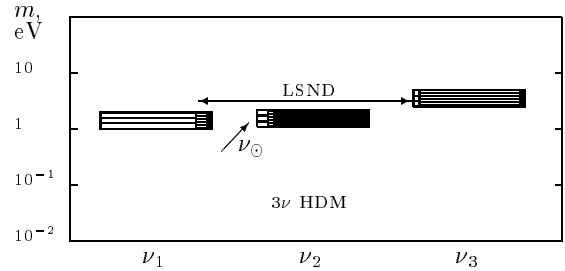,height=1.4in,angle=0}}
\caption[]{The same as in fig.1. Scenario without explanation of 
the atmospheric neutrino deficit.  
\label{fig:6}}
\end{figure}
According to \cite{fpq} (fig. 6) all neutrinos are in the eV - range, 
first two neutrinos are strongly degenerate: 
$\dmot = \msun$,  whereas $\dmo3 = \mlsnd$. 
Mixing is small: the electron flavor dominates in \none, the  
tau flavor -- in \ntwo ,  and the muon flavor in the heaviest 
state \n3. Remarks:\\ 
\noindent
(i) Solar neutrino problem is solved by 
the \ne \ra \nt small mixing MSW solution.\\ 
(ii) All three neutrinos give comparable 
contributions to the HDM.\\ 
(iii) \nm - \ne oscillations can be  in the  range 
of sensitivity of the LSND/KARMEN.\\ 
(iv) The Majorana mass $m_{ee}$ is at the level of 
upper experimental bound.\\

Another version \cite{rs} is characterized by 
$m_1 \ll m_2 \approx m_3 \sim 1$ eV 
with $\dmt3 = \msun$. 
Heavy components \none and \n3 are strongly mixed in 
\ne and \nt  and the lightest state has mainly the 
muon flavor (inverse hierarchy). 
(In contrast with 
scheme of sect. 3.3 , now the splitting between  
heavy states explain the solar neutrino problem.) 
Comments:\\ 
\noindent
(i) \ne \ra \nt conversion gives 
large mixing MSW solution to the solar neutrino problem.\\ 
(ii) The mass $m_{ee}$ can be at the experimental bound, 
although the cancelation is possible. \\
(iii)  One expects strong \bnum \ra \bnue 
conversion in the supernova, 
which is disfavored by SN87A data.\\ 
(iv) scenario supplies two component HDM and 
explanation of the LSND result. 

\subsection{``Standard" scenario}  

The scenario is characterized by strong mass hierarchy 
$m_1 \ll m_2 \ll m_3$ and 
weak mixing (fig.7).  
Basic features are:\\ 
\noindent
(i) $m_3 = m_{HDM}$,  so that 
\n3 forms the HDM.\\ 
(ii) Second mass, $m_2$,  is in the range
$m_2 = (2 - 3)\cdot 10^{-3} {\rm~eV}$ ,  
and  the 
$\nu_e \to \nu_{\mu}$ 
resonance  conversion 
solves the solar neutrino problem. \\ 
(iii) There is no solution of the atmospheric neutrino problem.\\ 
(iv) The depth of $\bar{\nu}_{\mu} - \bar{\nu}_e$  
oscillations with $\Delta m^2 \approx m_3^2$  equals
$4|U_{3\mu}|^2 |U_{3e}|^2$. Existing
experimental bounds on these matrix elements 
give the upper bound on this depth:
$ < 10^{-3}$ 
which is too small to explain the LSND result.\\
(v) Parameters of \nm - \nt oscillations can be in the 
range of sensitivity of the CHORUS/NOMAD. \\
\begin{figure}
\centerline{
\psfig{figure=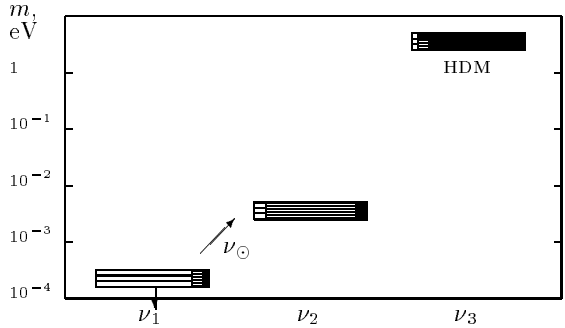,height=1.7in,angle=0}}
\caption[]{The same as in fig.1. The ``standard scenario. 
\label{fig:7}}
\end{figure}
There is a number of attractive features of this scenario: 
It naturally follows from 
the see-saw mechanism with 
Dirac mass matrix $m^D_{\nu} \sim m_{up}$ and 
the intermediate  mass scale for the Majorana mass matrix  of the 
RH neutrinos: $M_I \sim 10^{13}$ GeV.  
More precisely, for the eigenstates of this matrix 
one gets \cite{asmi}  
\begin{eqnarray}
M_2 & \sim & \left( 2 - 4 \right) \cdot 10^{10}~~ {\rm GeV}  
\nonumber \\  
M_3 & \sim & \left( 4 - 8\right) \cdot 10^{12}~~ {\rm GeV} \ .
\label{inter} 
\end{eqnarray}
These values of masses are in agreement with
``linear" hierarchy: $M_2/M_3 \approx m_c/m_t$.

The decays of the RH neutrinos with mass
$10^{10} - 10^{12}$ GeV can
produce the lepton asymmetry of the Universe
which can be  transformed by sphalerons into the
baryon asymmetry~\cite{fuk}. 

The 
mixing angle desired for solution of the $\nu_{\odot}$ problem 
is consistent with expression 
\begin{equation}
\theta_{e\mu} = \sqrt{\frac{m_e}{m_{\mu}}} - e^{i \phi} \theta_{\nu},
\label{emangle} 
\end{equation}
where $m_e$ and  $m_{\mu}$ are the  masses of the electron and muon,
$\phi$ is a phase and $\theta_{\nu}$ is the angle which comes from
diagonalization of the neutrino mass matrix.
The relation  between the angles and the masses (\ref{emangle})
is similar to  relation in  quark sector.    
Such a possibility can be naturally realized in terms of the
see-saw mechanism.

\subsection{More neutrino states?}

Another  way to accommodate all the anomalies is to introduce
new neutrino state  which mixes  with active neutrinos 
(see e.g. \cite{pelt,ptv,juha}).
As follows from LEP bound on the
number of neutrino species 
this state should be sterile (singlet of SM symmetry group). 
Mixing of sterile and   active neutrinos  leads to oscillations and the 
oscillations  result in  production of sterile neutrinos in 
the Early Universe. 
Presence of the sterile component in  
the epoch $t \sim 1$ sec  could influenced the 
Primordial Nucleosynthesis. 
Several comments are in order. 

1.  At present a situation with  bound 
on the effective number of the 
neutrino species, $N_{\nu}$, is controversial. 
Depending on the abundance  of primordial deuterium 
one uses in the analysis the bound 
ranges from 
$N_{\nu} < 2.5$ \cite{cris} 
to 3.9 \cite{ncris}. Certain model of evolution of the deuterium is used. 
A conservative analysis which does not rely on any model 
leads to $N_{\nu} < 4.5$ \cite{sar}. 
If  $N_{\nu} > 4$ is admitted then obviously 
there is no bound on oscillation parameters of the sterile 
neutrinos. 

2. Even if $N_{\nu} < 4$,   
strong mixing of the sterile and active neutrinos is not excluded.  
The bound can be avoided in presence of large enough 
($\Delta L \aprge 10^{-5}$)   
lepton asymmetry in the Universe,  
as it was discussed in sect. 2.5.

There are  bounds on 
oscillation parameters of sterile neutrinos 
from SN87A observations \cite{sn87a}.

Thus at present it seems possible  to introduce  
sterile neutrinos for explanations of different 
neutrino anomalies. 

\subsection{Rescue the standard scenario}

The atmospheric neutrino deficit is the problem for the standard 
scenario. To solve it 
one can  assume that an additional light singlet fermion 
exists with the mass $m \sim 0.1$ eV, which mixes mainly 
with muon neutrino, 
so that $\nm - \ns$ oscillations explain the data \cite{akhm}.
In this case one arrives at the scheme (fig. 8) \cite{js}. 
Production of \ns singlets  in the Early Universe 
can be suppressed (if needed) by
generation of the lepton asymmetry \cite{foot} in
the $\nt - \ns$ and $\bar{\nu}_{\tau} - \bar{\nu}_s$ oscillations 
\cite{thom}.
\begin{figure}
\centerline{
\psfig{figure=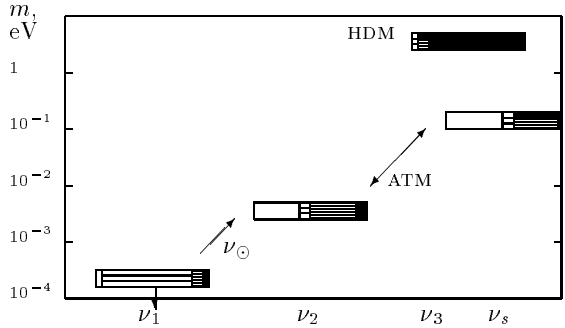,height=1.7in,angle=0}}
\caption[]{The same as in fig.1. Scenario with
sterile neutrino for atmospheric neutrino problem. 
The admixtures of sterile 
component are shown by white regions in boxes. 
\label{fig:8}}
\end{figure}
The presence of large admixture of the sterile component 
in \ntwo  influences resonance conversion of solar \ne, 
and also can modify the \nm - \nt oscillations \cite{js}.


\subsection{The safest  possibility ?}

Even without lepton asymmetry 
strong nucleosynthesis bound  is satisfied,  if \ns 
has the parameters of the solar neutrino problem.
In this 
scenario~\cite{cald,pelt} (fig. 9) 
$m_1  < m_S \ll m_2 \approx m_3$ . Remarks \\
\noindent
(i) Sterile neutrino has the mass $m_S \sim (2 - 3) \cdot 10^{-3}$ eV
and mixes with $\nu_e$, so that the resonance conversion
$\nu_e$ \ra $\nu_s$ solves the solar neutrino problem;\\
(ii) Masses of  ${\nu}_{\mu}$ and  ${\nu}_{\tau}$
are in the range 2 - 3 eV,  they supply
the  hot component of the DM; \\
(iii)  $\nu_{\mu}$ and  ${\nu}_{\tau}$
form the pseudo Dirac neutrino with large (maximal) mixing and
the oscillations ${\nu}_{\mu} - {\nu}_{\tau}$
explain the atmospheric neutrino problem;\\
(iv) The $\bar{\nu}_{\mu} - \bar{\nu}_e$ mixing can be strong enough
to explain the LSND result.\\ 
(v) No effect is  expected for \nm - \nt oscillations in  
CHORUS/NOMAD as well as in future searches of the \bdbd .  
\begin{figure}
\centerline{
\psfig{figure=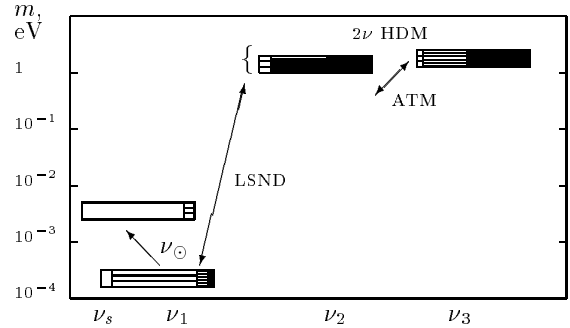,height=1.7in,angle=0}}
\caption[]{The same as in fig.1. 
Scenario with
sterile neutrino for  explanation of 
the solar neutrino deficit.  
The admixtures of sterile 
component are shown by white regions in boxes. 
\label{fig:9}}
\end{figure}

\section{On the models of neutrino mass}

\subsection{Predicting neutrino mass} 

Majority of  attempts to predict neutrino masses are reduced to 
establishing relations between quarks and leptons. 
Then   known parameters in  quark sector  
are used as an input to make some conclusions on 
mass and mixing in lepton sector. 

The see-saw mechanism allows one to realize the quark - lepton symmetry 
most completely. 
To make the predictions one should fix the Dirac mass matrix of 
neutrinos,  
$m^{D}_{\nu}$, 
as well as the Majorana mass matrix of the right handed  
components,  
$M_R$. 
Usually the direct Majorana masses  
of the left handed components are neglected. 
To find $m^{D}_{\nu}$, 
one can use  GUT relation, {\it e.g.} $m^D_{\nu} = m^{up}$ at GUT scale. 
For  $M_R$ different   ansatze \cite{anr} were  suggested. 
Also    
minimality of the Higgs sector can be postulated \cite{bamoh}
which allows one to get some 
relation between structure of $M_R$ and quark mass matrices. 
The pattern of   masses and mixing of the light neutrinos  
strongly depends on structure of $M_R$,  so that 
even for fixed
$m^{D}_{\nu}$, practically any scenario  can be 
realized.

Relations between  quarks and leptons 
can be based also  on 
certain horizontal symmetries.

Recent attempts to predict neutrino masses are based on 
\begin{itemize}
\item GUT models with $SO_{10}$ symmetry,
 
\item Models with anomalous $U(1)$ symmetry,  

\item SUSY Models with R-parity violation,  

\item Models with radiative neutrino mass generation. 
\end{itemize}

Also one can introduce  some ansatze for the 
quark and lepton matrices.

\subsection{An ansatz for large lepton mixing}

It is postulated \cite{frit} that fermion mass matrices have the following 
structure in certain basis (the scale is not specified) 
\be 
M_i = c_i M^{dem} + \Delta M_i^{diag}
\label{ansatz}
\ee 
$i = u,d,l,$ $\nu$, where 
\begin{equation} 
M^{dem} = 
M_0 \left(\begin{array}{ccc}
1  & 1  &  1   \\
1  & 1  &  1   \\
1  & 1  &  1   
\end{array}\right)
\label{dem}
\end{equation}
is the democratic matrix,  and 
\begin{equation} 
\Delta M^{diag} = 
\left(\begin{array}{ccc}
\delta  & 0  &  0   \\
0  & \rho  & 0    \\
0  &  0  &  \epsilon    
\end{array}\right) ~~~~
\label{diag}
\end{equation}
with 
$ \delta, \rho, \epsilon \ll M_0$. 
It is assumed that parameters $c_i$ are proportional 
to electric charges of fermions: 
\be
c_i \propto Q_i ~. 
\ee
For quarks and charged  leptons first term in 
(\ref{ansatz}) dominates leading to  big mass 
in one generation and big mixing angles which diagonalize 
 matrices. 
As the result of two similar rotations for the 
upper and for the down quarks, the mixing in quark sector is small. 
The situation in lepton sector is different. 
For neutrinos: $c_i = 0$, and therefore the neutrino mass matrix 
is diagonal: 
\be
M_{\nu} = \Delta M^{diag} ~. 
\ee
The lepton mixing follows from  diagonalization of the charge lepton mass 
matrix and since $M_l \approx M^{diag}$, the mixing 
in leptonic sector is automatically large. 
In a sense large mixing in lepton sector is  
related to  smallness of the neutrino mass.

All three neutrinos are strongly mixed. 
If $\delta \approx \rho \approx \epsilon \sim 1$ eV  
and splitting is very small: 
($\delta^2 - \rho^2) \sim 10^{-10}$ \ev and  
($\rho^2 - \epsilon^2) \sim 10^{-2}$ \ev ,  one can explain   
the solar 
neutrino data  via just-so oscillations,  the atmospheric 
neutrino deficit. and presence of the HDM.  

However there is no understanding why 
$\Delta M^{diag}_{\nu} \ll \Delta M^{diag}_{u,d,l}$.

\subsection{SO(10) model}

The (supersymmetric) model  \cite{chou} is based on 
$G = SO(10)\times \Delta (48) 
\times U(1)$ symmetry, where 
dihedral group 
 $\Delta (48)$  (subgroup of $SU(3)$) 
was used as the family symmetry 
\footnote{Previously dihedral group $\Delta (75) $ 
was suggested in \cite{kaplan}}. Fermions are in (16,3) 
representation of $SO(10)\times \Delta (48)$. The fermion masses are 
generated by Froggatt-Nielsen mechanism and for this a number of new 
chiral superfield is introduced. The symmetry is broken 
to the SM symmetry in three steps: 
$G$ \ra $SO(10)$ \ra $SU(5)$ \ra $SM$ at mass scales $M_P$, $v_{10}$, and 
$v_{5}$ 
correspondingly. All fermion are predicted in terms of 4 continuous 
parameters: 
$v_5/M_P$, 
$v_{10}/M_P$, the ratio of MSSM VEV: $\tan \beta$,  and 
universal Yukawa coupling $\lambda$. An additional 
$U(1)$ symmetry is also used to get desired structure 
of the mass matrices. 
The  neutrino Dirac mass matrix has the following hierarchical 
structure of the elements: 
$ 
m_{33} \gg   m_{12} \gg  m_{23} \aprge  m_{22},      
$ 
for  the Majorana mass matrix one gets: 
$
M_{22} \gg M_{13} \gg M_{33}
$,  
(the matrices are symmetric and all other elements are zero).    
This leads via the see-saw mechanism to the pattern of the light 
masses with 
$
m_2 \approx m_3 \gg m_1
$. 
Also additional sterile neutrino is introduced to explain 
the solar neutrino problem, thus the model 
reproduces  the pattern 
discussed in sect. 3.9.  
 
It should be stressed however that the pattern is the 
result of {\it ad hoc} introduction 
of the large number of new supermultiplets and special 
$U(1)$- charge prescription. In fact, these $U(1)$ charges 
should be considered as new free parameters, 
so that high predictivity becomes not so impressive.

\subsection{Neutrino-neutralino mixing}

This is low scale   realization of 
the see-saw mechanism. 
The neutrino mass equals  
$m_{\nu} \sim m_{\nu N}^2/ m_{N}$,  
where  
$ m_{\nu N}$ is the mixing mass term, and  
$ m_{N}$ is the typical neutralino mass. 
Mixing of neutrinos and neutralinos implies 
violation of the $R$-parity. It was realized recently that 
Hall - Suzuki model \cite{hall} endowed by  the universality 
of some  SUSY breaking mass terms leads 
naturally to  
$m_{\nu N} \ll  m_{N}$,  and therefore to smallness of the neutrino 
mass~\cite{pol,hempf,asfv,nilles}. 

In terms of the MSSM multiplets,   
the superpotential of the model at GUT scale is  
\be
W = \mu_i L_i H_2 + h_b Q_3 L_0 D^c_3   
\label{super}
\ee
($i = 0, 1, 2, 3$), where  $H_2$ is the Higgs doublet,  
and $L_0 \equiv H_1$  
is defined as the only  component which 
has Yukawa 
couplings at GUT scale;  
(we took into account the Yukawa 
couplings of the third generation only).   
$Q_3$ is the quark doublet, $D^c_3$ is the    
superfield with the RH quark component. 

The model implies that  the $R$-parity 
is broken by dimension three (and less) operators only. 

Basic feature of the superpotential is that 
only one component of the quartet, $L_0$ 
has the Yukawa couplings. This can be related to  
$R$-symmetry \cite{nilles}.  The fields $L_0$ and  $L_a$  (a = 1,2,3)
may have different $R$-charges: 
e.g.  $R(L_0, H_2) = 2$,  whereas 
$R(Q, U^c, D^c, L_a ) = 0$. In this case 
the $R$-parity breaking Yukawa couplings are suppressed.  
Moreover, the $\mu$-terms can be generated by nonrenormalizable 
interactions with new fields $z_i$.  
The  $R$-symmetry is 
broken spontaneously 
by the VEV of 
these  field $z_i$: 
$\langle z_i   \rangle  \ll M_P$, and the 
$\mu_i$ parameters of the superpotential may have 
the hierarchy determined by 
$\langle z_i   \rangle ^n / M_P^{n-1}$, where $n$  
is fixed by the $R$-charge of $z_i$ \cite{nilles}. 

It is assumed 
(here we will follow discussion in \cite{asfv})   
that soft SUSY breaking terms 
for $L_i$ are universal at,  {\it e.g.},  GUT scale: 
\be
V = B\cdot \mu_i L_i H_2 + m_0^2 
|\tilde L_i|^2 + ...
\label{potential}
\ee
Due to the universality one can diagonalize  the 
$\mu$ term in the superpotential,  and simultaneously  
in the potential (\ref{potential}),  
by rotation $L_i$ \ra  $L_i'$: 
\be
\mu_i \tilde{L}_i H_2 \rightarrow  \mu L_0' H_2 ~. 
\label{rot}
\ee 
(This rotation generates simultaneously the $R$-parity violating 
Yukawa couplings).    
There is no terms like $\tilde{L}_a' H_2$ (a = 1,2,3) at GUT scale. 
These terms however appear at the electroweak scale due to 
the renormalization group effect.  
Indeed, Yukawa coupling  (\ref{super}) 
distinguishes different components 
of $L_i$ and this  leads to different renormalization 
of  terms  with $\tilde{L}_0$ and $\tilde{L}_a$ 
in (\ref{potential}). 
The universality turns 
out to be broken, and the rotation  (\ref{rot}) 
will not diagonalize the  potential. 
We get after rotation (\ref{rot})
the mixing term  
\be
\frac{\mu_i}{\mu}  \times
\left[
\delta m^2\ {L'_0}^* + \delta B\cdot \mu\ H_2
\right] 
\tilde L'_i +
{\rm h.c.}
\label{part}
\ee 
where $\delta m^2$ and $\delta B$   
describe the renormalization 
group effect. 
After electroweak symmetry breaking
the mixing terms (\ref{part}) (linear in $\tilde{L}'$), 
together with soft symmetry breaking masses,
induce a VEV of  ``sneutrino" (neutral component 
of the doublet in $\tilde L_i' $) 
of the order:
\be
\langle \tilde {\nu} \rangle \approx   
v \  \frac{\mu_i}{\mu} \times
\left(\frac{\delta m^2}{m_{0}^2}\ \cos\beta +
\frac{\delta B\cdot \mu}{\; m_{0}^2}\ \sin\beta \right)~,
\label{sneutrinovev}   
\ee 
here $v$ is the electroweak scale. 
The VEV of sneutrino leads via the gauge coupling to the 
neutrino-gaugino mixing: 
$m_{\nu N} \sim g \langle \tilde {\nu} \rangle $. In turn the 
see-saw mechanism results in the mass 

\begin{eqnarray}
m_{\nu} & \approx & 
\frac{g_1^2+ g_2^2}{2}\ \frac{\langle \tilde{\nu} \rangle^2}{M_{\tilde Z}}
\nonumber \\
& \approx &    
9 \frac{m_Z^2}{M_{\tilde Z}} 
\left(\frac{\mu_i}{\mu}\right)^2
\left[ 
\frac{h_B^2}{16 \pi^2} \log \frac{M_{GU}^2}{m_W^2}
\right]^2~.  
\label{mass}
\end{eqnarray}
This contribution to  neutrino mass is
typically larger
than the one produced by the loop-diagram stipulated by 
$R$-parity violating Yukawa couplings.   
For $\mu_i \sim \mu$ and large $\tan \beta$ ($h_B \sim 1$) 
we find $m_{\nu} \sim O(10$ MeV). This neutrino can be identified 
with \nt.

There are several possibilities to get much smaller 
mass. 
For small $\tan \beta (\sim 1)$ the Yukawa coupling is 
small and the $m_{\nu}$   is of the order 10 eV. 
Also the mass can be suppressed if 
there is the hierarchy of  
$\mu_i$: $\mu_i/\mu \ll 1$ .   
For  
$\mu_i/\mu  \sim M_{GUT}/M_{Pl}$ :   
$m_{\nu} \sim 10 $ eV even for large $\tan \beta$~\cite{hempf}. 

Another  possibility  
for suppression of $m_{\nu}$  is a  
cancelation between the two terms in (\ref{sneutrinovev}). 
If there is no cancellation,  the neutrino mass 
turns out to be related to the 
$R$-parity violating Yukawa coupling generated by rotation 
(\ref{rot}): $\lambda^4 \approx C m_{\nu} m_{\tilde{Z}} 
(\mu_i/\mu)^2$, where $C$ is known constant \cite{asfv}. 
Thus certain relation between the probabilities of $R$-parity 
violating processes (due to $\lambda$) and neutrino mass 
gives  signature of this mechanism.

In the case of three generations only \cite{pol,hempf}
one neutrino acquires the mass due 
to neutrino-neutralino mixing. 
Loop corrections induced by $R$-parity violating couplings 
make all neutrinos massive. 
In certain region of parameters one can explain 
solar neutrino problem and supply HDM (i.e. reproduce 
the standard scenario).  Also simultaneous solution 
of the solar and  atmospheric neutrino problems is possible 
\cite{hempf}.

\subsection{Models with anomalous $U(1)_A$ symmetry}

Masses of neutrinos are generated by the see-saw mechanism. 
Structure of the neutrino mass matrices is determined by 
$U(1)_A$ charges of neutrinos \cite{drei,pap,binet,leon}.  
Relation between the neutrino  and the quark mass matrices 
is established via the charges (rather than immediately,  as in 
the simplest GUT theories).  
It is assumed that charges of neutrinos coincide with charges of 
(electrically charged) leptons:  
\be
Q(\nu_{l L}) = 
Q(\nu_{l L}^c) = 
Q(l) = Q_l ~. 
\label{charges}
\ee 
Masses are generated {\it a la} Froggatt-Nielsen mechanism,  and  
elements of  the mass matrix appear as  
\be
m_{ij} \sim m_0 \left(\frac{\theta}{M}\right)^{|Q_i + Q_j + Q_H|}~, 
\ee 
where  $\theta$  is the VEV 
of singlet field with unit  $U(1)_A$ charge and $M$ 
is the mass scale of new heavy scalar fields. 
There are  different mass parameters  
for the upper, $M_2$, and down, $M_1$, 
fermions. The equality of charges (\ref{charges}) 
leads to the following relation between the 
neutrino Dirac mass matrix and the mass matrix 
of the charged leptons: 
\be
m^{D}_{\nu} \sim  
\tan \beta \cdot m_l(M_1 \rightarrow M_2)~. 
\label{dirac}
\ee 
(In  the leptonic matrix one should substitute 
$M_1$ \ra $M_2$). 

The Majorana mass matrix of the RH neutrino components is generated by  
coupling with new singlet field $\Sigma$: 
\be 
M_R \sim  \langle \Sigma \rangle \cdot 
\left(\frac{\theta}{M_3}\right)^{|Q_i + Q_j + Q_{\Sigma}|}~. 
\ee
Depending on the charge of the $\Sigma$, 
$Q_{\Sigma}$, (which is, in fact, unknown) one can get different 
structures of 
$M_R$ and eventually of the mass matrix of light neutrinos.  
The $\Sigma$ can appear as the composite operator: 
\be
\Sigma = \frac{s \cdot s}{M_P}~. 
\ee
Here $M_P$ is the Planck mass. 
For $\langle s \rangle \sim 10^{16}$ GeV one gets  
$\langle \Sigma  \rangle \sim 10^{13}$ GeV. 

Depending on charge prescription (especially for $\Sigma$) 
one can accommodate the solar neutrino data  and HDM 
(sect. 3.6, fig. 7),  or the solar 
and atmospheric neutrino problems (sect. 3.4, fig.5).  

\subsection{Zee model revisited} 

Zee model~\cite{zee} 
includes  the charged scalar field
$h$, being a  singlet  of the $SU(2)$,  and  two
doublets of the Higgs bosons. 
In virtue of the gauge symmetry 
the singlet  $h$  has antisymmetric (in flavor) couplings to lepton doublets
$L_{lL} \equiv (\nu_l, l^-)$, ($ l = e, \mu, \tau$)
\begin{equation}
{\cal L}_{Zee} =
f_{\ell \ell'} L_{\ell L}^T i\tau_2  L_{\ell'L} h \
+ h.c. ,
\label{lagr}
\end{equation}
\noindent
$f_{\ell \ell'} = -  f_{\ell' \ell}$.   
Neutrino mass is generated  in one
loop. Neutrino mass terms are proportional to masses of 
the charge leptons 
squared. As the consequence of the  antisymmetry of the couplings 
and hierarchy  of charge leptons masses, the 
Zee model gives very distinctive pattern of
neutrino masses and mixing \cite{wo} .
For not too strong  hierarchy of  couplings $f_{\ell \ell'}$
the two heavy neutrinos,
$\nu_2$,  $\nu_3$,  are
degenerate  and mix in $\nu_{\mu}$ and $\nu_{\tau}$
almost maximally. The first neutrino $\nu_1$ practically coincides
with $\nu_e$ and has much smaller mass:
$  
m_1$   $ \ll $  $m_2$   $ \approx$  $ m_3 .
$
This pattern  coincides  with the one (fig. 2) needed  for  
a solution of  the atmospheric neutrino problem
by $\nu_{\mu} \leftrightarrow \nu_{\tau}$ oscillations
and for  existence of the two component hot dark matter
in the Universe \cite{stao}. 
Furthermore, the oscillations
$\bar\nu_{\mu} \leftrightarrow \bar\nu_e$
can be in the range of sensitivity of   
KARMEN/LSND experiments \cite{wolfz,tans}.
Thus the model reproduces the pattern discussed in sect. 3.3, 3.7.
The analysis shows \cite{tans} that 
scenario 
implies large values and inverse flavor  hierarchy of couplings
of the Zee boson with fermions:
$f_{e \tau} \ll f_{\mu \tau} \leq f_{e \mu} \sim 0.1$.

Main signatures of scenario  are: strongly suppressed
signal of $\nu_{\mu} \leftrightarrow \nu_{\tau}$ oscillations 
in CHORUS/NOMAD, so that positive result
from these experiments will  rule out the scenario;
possibility of observation of
$\nu_e \rightarrow \nu_{\tau}$ oscillations by CHORUS/NOMAD;
corrections to the
muon decay and  neutrino-electron scattering at the level
of  experimental errors;
branching ratio $B(\mu \rightarrow e \gamma)$  bigger than $10^{-13}$.
The solar neutrino problem can be solved
by introduction of additional very light singlet fermion
without appreciable modification of the
active neutrino pattern. 

\subsection{Sterile neutrinos: window to the  hidden world ?} 

Common wisdom is that existence of light sterile neutrinos 
is unnatural.  Indeed,  
introducing sterile neutrino, \ns,  one encounters several questions: 
What is the origin of this neutrino? 
How it mixes with usual neutrinos? 
What protects the mass of \ns  and makes it  small ? 
Therefore  discovery  of  sterile 
neutrino will mean something very non trivial. 
In fact,  forthcoming solar neutrino experiments, 
as well as the atmospheric neutrino experiments and 
long base line experiments will be able to establish,  
if solar or/and  atmospheric neutrinos are converted into 
sterile neutrinos.

What could be behind this discovery? 
There are several studies of this question recently.

1. Immediate candidate for \ns is the RH neutrino component.
However in this case  the see-saw mechanism does not operate.

2. Sterile neutrino   could be the component of  multiplet 
of some  extended 
gauge symmetry - like  $SO(10)$-singlet from  27-plet of $E_6$
\cite{ma}. The mass of the \ns is generated by 
separate see-saw   mechanism and its value  is protected by 
$U(1)$ symmetry which is embedded in $E_6$ and broken 
 at low scale.

3. In~\cite{zur} it was suggested that \ns is the mirror neutrino
from the mirror standard model. The mass of \ns is generated 
by the see-saw mechanism in the mirror world 
which, however, has the electroweak symmetry 
breaking scale 
$\langle H_M \rangle $
about two orders of magnitude bigger than in usual world. 
(Here  $H_M$ is the mirror Higgs doublet.)  
Generalizing (1) we get 
$
m_{s} = 
\langle H_M \rangle^2 /M_P
$ .  
Mixing of usual neutrinos  with the mirror one 
proceeds via the gravitational interactions 
\be
\frac{1}{M_P} \bar{L} H \bar{L}_M  H_M  + h.c.~, 
\ee 
where $L_M$  
is  the mirror lepton doublet.  
Therefore the mixing angle 
is determined essentially by the ratio of VEV: 
$\langle H  \rangle / \langle H_M \rangle $. 


4. The origin and properties of \ns can be related to SUSY. 
A number of singlet superfields was
introduced for different purposes: to generate $\mu$ term,
to realize PQ-symmetry breaking, to break spontaneously the lepton
number, {\it etc.}.   String theory typically  supplies a number of 
singlets. Fermionic components of these superfield could be
identified with desired sterile neutrino. 

It was shown  in \cite{cjs1} that masses and mixing of \ns can be 
protected by $R$-symmetry.

5. Another possibility is that \ns is the 
would be Nambu-Goldstone fermion \cite{cjs2}: 
the superpartner of the Nambu-Goldstone boson 
which appears as the result of spontaneous violation 
of some $U(1)$ global symmetry like Peccei-Quinn 
symmetry or lepton number symmetry etc. 
(i.e. \ns is the axino,  or majorino ....). 
General problem is that susy breaking generates 
typically the mass of \ns of the order the gravitino mass 
and further suppression is needed. One can use here 
the ideas of  non-scale supergravity, or possibly,  
gauge mediated SUSY breaking.

6. Sterile neutrino as modulino? 
Suppose that there is a singlet $S = \ns$ 
which is  massless
in the supersymmetric limit and couples with  
observable sector  
via the gravitational interactions. The  
mass and  effective interactions are 
induced when supersymmetry is broken.  
For some reasons ({\it e.g.} relataed to cancellation of the 
cosmological constant) $S$ may not acquire the mass in the order 
$m_{3/2}$. Then  natural scale of  mass of $S$ is  
\begin{equation}
m_S\sim \frac{m_{3/2}^2}{M_{P}} ~.  
\label{ms1}
\end{equation}
The mixing of $S$ with active neutrinos involves electroweak symmetry
breaking. The simplest appropriate effective operator is 
($m_{3/2}/M_{P}) LSH$.  It  generates the mixing mass  
parameter 
\begin{equation}
\bar m\sim \frac{m_{3/2}\left\langle H\right\rangle }{M_{P}}~.   
\label{mbar1}
\end{equation}
For small electron neutrino mass $m_{\nu _e}\ll m_S$ the $\nu _e-\nu _S$ 
mixing angle $\theta_{es} $ is of the order
\begin{equation}
\theta_{es} \sim \frac{\left\langle H\right\rangle }{m_{3/2}}~.   
\label{theta1}
\end{equation}
For $M_{P}\sim 2\times 10^{18}$
GeV and $m_{3/2}\sim 10^3$ GeV  
one gets $m_S$ and $\bar m$ 
\be
m_S \sim  10^{-3} {\rm eV}~, ~~~
 \bar m  \sim  2 \cdot 10^{-4}\rm{eV}
\ee
precisely in the range desired for 
a solution of the solar neutrino problem 
via resonance 
conversion $\nu _e\rightarrow S$ in the sun. 
Moreover, varying the
parameters (constants of the order 1) and taking into 
account the renormalization group effect 
it is easy to achieve both small and large mixing solutions to
the problem.

Fermion $S$ can also mix with the other neutrino species. If the
coupling of $S$ with fermion  generations is universal; 
{\it i.e.}  $\bar m_i$ are
the same (or of the same order) 
for all generations,  then $S$-mixing with
$\nu _\mu $ and $\nu _\tau $ 
are naturally suppressed as the mixing angles  
behave as $\theta _i\sim \bar m/m_i$. 
For instance,  taking $m_2\sim 10^{-1}$ eV 
and  $m_3\sim 1$ eV 
we get $\sin ^22\theta _{S\mu }\sim 10^{-5}$ 
and $\sin ^22\theta _{S\tau }\sim 10^{-7}$. 
Thus the lightest neutrino has
naturally the biggest mixing with \ns.

The desired properties  of $S$ could be realized for some 
fields in hidden sector, and probably for fermionic 
components of some moduli fields\cite{benak} . 

\section{Conclusion}

1. New effects of the  neutrino refraction 
in media have been  considered recently which may have important 
impact on pattern of neutrino masses and mixing.  

Neutrino conversion  in polarized and magnetized media  
opens  new possibility in explanation 
of peculiar velocities of pulsars. This implies 
$m_{\nu} \aprge 100$ eV. 

Large  leptonic asymmetry in the 
Early Universe due to oscillation into sterile neutrinos  
may have serious impact on primordial 
nucleosynthesis and the nucleosynthesis  bounds on 
neutrino parameters

Modification of long range  forces  stipulated by the neutrino exchange 
in dense medium allows one to resolve the  
energy paradox in compact stellar objects (neutrons stars, 
white dwarfs etc..) \\

\noindent 
2. Several possible patterns (scenarios) 
of neutrino masses and mixing were elaborated on the basis 
of  present  neutrino data  (hints and bounds). 
This allows one to check a consistency of different 
positive results  
and gives a guideline for further studies. 

The data indicate  that  
structure of the mass spectrum and lepton mixing may differ 
strongly from those in quark sector. 
In particular, spectrum may show complete degeneracy, 
pseudo Dirac structure, or even inverse hierarchy. 
The mixing can be large or even maximal. 
New sterile states may exist which mix with active neutrinos. 

Different scenarios have rather distinctive predictions and 
forthcoming experiments (SK, SNO, CHOOZ CHORUS/NOMAD, NEMO ....) 
will be able to discriminate among them.\\ 

\noindent
3. Neutrinos may have several different sources of mass:  
usual  see-saw contribution,  radiative effects, 
mixing with neutralinos (in models with 
$R$-parity violation). Structure of the  mass 
matrices can be related to supersymmetry and  $R$-symmetry.  
The neutrino mass and mixing can have a connection 
to quark-lepton symmetry, GUT, to new mass scales 
and new symmetries. 

However it will be difficult to identify mechanism 
of neutrino mass generation just from neutrino data 
(even if in future 
we will know neutrino parameters with good precision). 
As an illustration: two different models 
discussed in sect.4  radiative Zee model and 
GUT $SO(10)$ with horizontal symmetry  lead to 
precisely the same pattern in lepton sector. 
To identify the mechanism one will need an information about  
other elements of models: {\it e.g.} the discovery  of 
proton decay, 
processes with R-parity violation,  Zee singlet {\it  etc.  },  will 
clarify many points. \\

{\bf Acknowledgments}: 
I am grateful to  A. Joshipura, and F. Vissani  
for fruitful discussions.

\section*{Questions}
\noindent{\it D. R. O. Morrison, CERN}

Dr. Smirnov has mentioned a paper that I did not have time to submit 
as I was working on molecular genetics. However I did send him a paper written 
earlier this year which raises worries about 3 things- Errors, the Sun's  
luminosity and motion inside the Sun.

1. ERRORS: We seem to agree there is a problem with the very different 
errors of different SSM's. 

2. LUMINOSITY: What we measure is what we see on the surface of the Sun over 
the last few years. But what we need 
is an average over the last few million  
years as the time for thermal information to travel from the core to the 
surface is between one and ten million years (Douglas Gough's estimate).
The latest satellite measurements 
show the luminosity follows the sunspot cycle.
If we were living near the year 1700, the luminosity would have been quite 
different as the earth's temperature was much lower - 
in London people had 
fairs with bonfires on the ice on the Thames - and there were no sunspots 
between 1650 and 1710. 
Similarly there were few sunspots about 1400 when there 
was another cold spell whereas near 1200, there was a hot period with extra 
sunspots. 
   In other words, the surface of the Sun changes in ways not included in the 
SSM which does not consider sunspots nor variation of the apparent luminosity. 
  Going back further, for many million years, the sea level was much higher 
indicating that the luminosity was much greater. For example when the  
dinosaurs were extinguished, the sea level was consistently about 200 metres  
higher than now and half of the present land surface was under water.
  We do not have a good measurement of the 
luminosity over a suitably long 
time period and hence the error on the 
luminosity should be greatly increased.  

3. INTERNAL MOTION: There are three pieces of evidence. 
Initially the Sun was  
a T Tauri star - very bright and rotating quickly. 
Standard Solar Models 
cannot slow this rotation to zero, 
so one expects a differential rotation 
even to the core of the Sun. This is supported by helioseismological 
measurements which show that the rotation at the poles and at the equator is
different down to 0.2 of the Sun's radius. 
   Helium-3 has an unusual 
distribution being sharply peaked at a radius 
of 0.3. Calculations by Wick Haxton have shown that a motion of only 
700 metres per year, is enough to cause 
this Helium-3 to move and to be burnt 
thus changing the temperature of the Sun's core appreciably.
  Lithium-7 has a measured abundance which is less than one hundredth of that  
predicted by the SSM. Also looking at 
other stars, the Boesgaard dip is not 
explained by the SSM. Sylvie Vauclair et al. 
have explained this by meridional
motion inside the Sun. 
However they cut the motion at a radius higher than 0.3 
and hence do not allow any Helium-3 movement, 
and so find little change in the 
neutrino flux. Without this cut which seems in contradiction to the 
helioseismological results which show effects down 
to at least 0.2 radius, the
neutrino flux would have been changed.

\vskip 12pt
\noindent{\it A.Yu. S.:}\\
You said so many things that I forget what they were.
 
\vskip 12pt
\noindent{\it D. R. O. Morrison}:\\   
Errors, Luminosity, Internal motion.

\vskip 12pt
\noindent{\it A.Yu. S.:}\\
1. At present the solar neutrino problem can be formulated 
practically without reference to a specific standard solar model. 
The problem can be formulated as discrepancy between different experimental 
results, and in this connection more  relevant question is how 
reliable are the experimental results.  

2. Possible variations of the Solar luminosity are certainly 
much smaller than those which could correspond to 
depletion of the Gallium result by factor 2. 

3.  W. Haxton has suggested   unusual mixing of elements.  
It involves fast filamental flow of matter from the layers
with maximal concentration of $^3He$ downward, and
slow restoring flow upward. This leads to enhancement of
the pp-I branch and therefore suppression of the
$^7 Be$ neutrino flux. 
However, (i) the 
suppression  achieved by  mixing  
is not enough  for good description of data.    
(ii) In fact,  no consistent solar model has been elaborated 
which incorporates the mixing, 
and it is unclear what is a  feedback of the  
mixing on other observable characteristics of the Sun. 
(iii) It has been shown by Bahcall and 
collaborators, that  mixing siggested by Haxton strongly 
contradicts the helioseismology data.

Results obtained by  S. Vauclair {\it et al}   
show that solution of the Lithium-7 problem 
has no serious impact on solar neutrino fluxes if one takes 
into account the helioseismological data.  
The rotation-induced mixing of elements below the
connective zone is introduced; 
the agreement with the helioseismological data 
imply that mixing should terminate 
below $R < 0.4 R_{\odot}$.  This is enough 
to solve the 
lithium problem, but this is not enough to change 
appreciably the neutrino fluxes 
which are generated in deeper layers.  

In paper presented at Rencontres de Blois 
Vauclair {\it et al.},  add 
some mixing in the central 
regions ($R \sim  (0.1 - 0.2) R_{\odot}$)   
to have  better description of 
the helioseismology data. 
It turns out that  the mixing should be
weak  and its effect on 
neutrino fluxes is of the order 20\% only.

\vfill\eject

\end{document}